\documentclass[twocolumn,showpacs,preprintnumbers,amsmath,amssymb]{revtex4}

\usepackage{graphicx}
\usepackage{dcolumn}
\usepackage{bm}

\begin{document}

\title{Optimal transport on supply-demand networks}
\author{Yu-Han Chen$^1$}
\author{Bing-Hong Wang$^{2,3}$}
\author{Li-Chao Zhao$^4$}
\author{Changsong Zhou$^1$}
\author{Tao Zhou$^{2,5}$}
\email{zhutou@ustc.edu}
\affiliation{$^1$Department of Physics, Hong Kong Baptist University, Kowloon Tong, Hong Kong \\
$^2$Department of Modern Physics, University of Science and
Technology of China, Hefei 230026, P. R. China\\
$^3$Research Center for Complex System Science, University of
Shanghai for Science and Technology, Shanghai 200093, People's
Republic of China \\
$^4$Department of Physics, Oklahoma State University, Stillwater, OK
74075, USA\\
$^5$Department of Physics, University of Fribourg, Chemin du Muse 3,
Fribourg 1700, Switzerland }

\date{\today}

\begin{abstract}
Previously, transport networks are usually treated as homogeneous
networks, that is, every node has the same function, simultaneously
providing and requiring resources. However, some real networks, such
as power grid and supply chain networks, show a far different
scenario in which the nodes are classified into two categories: the
supply nodes provide some kinds of services, while the demand nodes
require them. In this paper, we propose a general transport model
for those supply-demand networks, associated with a criterion to
quantify their transport capacities. In a supply-demand network with
heterogenous degree distribution, its transport capacity strongly
depends on the locations of supply nodes. We therefore design a
simulated annealing algorithm to find the optimal configuration of
supply nodes, which remarkably enhances the transport capacity, and
outperforms the degree target algorithm, the betweenness target
algorithm, and the greedy method. This work provides a start point
for systematically analyzing and optimizing transport dynamics on
supply-demand networks.
\end{abstract}


\maketitle

\section{Introduction}
Network transport has attracted increasing attention in recent years
(see the review articles \cite{Wang2007,Tadic2007} and the
references therein). Indeed, it describes a large number of natural
phenomena and technological processes, such as substance flow in a
metabolic network, power transmission in an electric network,
information propagation in the Internet, and so on. A matter of
prime importance is to make the transport processes more effective
and efficient, corresponding to maximizing the global capacity and
minimizing the average delivery time. Previous works addressed this
issue can be roughly classified into two categories: one concerns
the optimization/modification of underlying topology
\cite{Guimera2002,Cholvi2005,Zhang2007}, while the other focuses on
the design of highly efficient transport/routing protocols
\cite{Echenique2004,Yan2006,Danila2006,Sreenivasan2007,Zhou2008}.

A latent assumption in most previous works is that every node in a
transport network plays the role of a host, that is to say, every
node has the ability creating a certain kind of substance, energy or
information. However, the real world is far from this assumption.
For example, in an electric network
\cite{Carreras2002,Carreras2004}, there are two kinds of nodes,
power stations and transformer substations. The power is generated
in the former nodes, flowing to the latter ones, and then imported
to customers through them. Therefore, power stations behave as a
kind of suppliers, while the transformer substations are customers
holding demands. In some Internet serving systems, such as music
libraries (e.g., audioscrobbler.com, see Ref. \cite{Lambiotte2005}),
movie-sharing systems (e.g., Netflix.com, see Ref. \cite{Zhou2008b})
and on-line viewing site (e.g., YouTube.com, see Ref.
\cite{Crane2008}), all the resources are located in a few servers,
while other connected nodes, usually personal computers, only regale
themselves with those services. Those examples give rise to a
general concept of \emph{supply-demand network}, whose nodes are
classified into two categories: the supply nodes provide some kinds
of services, while the demand nodes play the role of customers.
Analysis of supply-demand networks has found its applications in
various real systems, ranging from the power grid
\cite{Liu1984,Bai2005} to supply chain networks
\cite{Cheng2006,Goh2007}.

In this paper, we propose a general model for the transport on a
supply-demand network, whose capacity is very sensitive to the
locations of supply nodes. By applying a simulated annealing
algorithm, we obtained the near optimal locations of supply nodes
subject to the maximal network transport capacity. The proposed
algorithm performs obviously better than the random selection,
degree targeted, betweenness targeted, and greedy methods.


\section{Model}
Considering a network consisted of $N$ nodes, which are classified
into two categories: One is called the \emph{supplier} that provides
a certain kind of service, the other is called the \emph{customer}
who requires this service. Here, the service is an abstract concept
and can stand for substance, energy, information, etc. For
simplicity, we use the language of the Internet, that is to say,
every customer need some information packets (resource), and only
the suppliers can generate those information packets. We assume the
demands are uniformly distributed, namely each customer needs a unit
number of packets (one can simply say one packet). For a given
customer, we suppose this packet is always sent by one of the
nearest suppliers. However, in general case, there are several
nearest suppliers and for each there are several shortest paths. In
the real implementation, one of those shortest paths should be
randomly picked, and the packet will follow this path from the
supplier to the customer. In the numerical calculation, to reduce
the fluctuation, if there are in parallel $k$ shortest paths from a
customer to the suppliers (generally, those paths aim to more than
one nearest suppliers), we assume the packet is divided into $k$
pieces, each goes through one shortest path and contributes $1/k$ to
the traffic load (see an illustration in Fig. 1).

\begin{figure}
\scalebox{1}[1]{\includegraphics{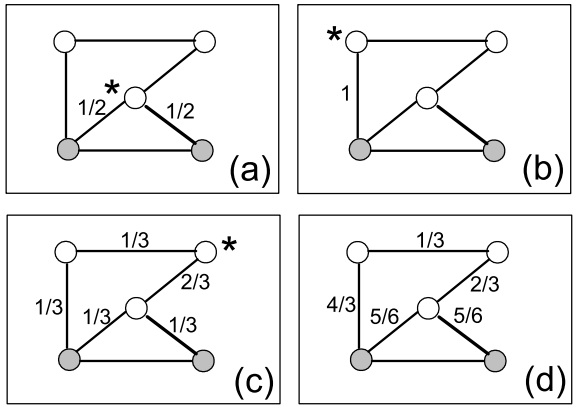}} \caption{Illustration
of the distribution of edge loads in a supply-demand network. The
gray solid and hollow circles denote supply nodes and demand nodes,
respectively. In each of panels (a), (b) and (c), the circle marked
by a star is the target demand node, and the resulting loads are
labeled besides corresponding edges. Integrating (a), (b) and (c),
the distribution of edge loads can be obtained, as shown in the
panel (d). Here, $L_{\texttt{max}}=4/3$.}
\end{figure}

If the bandwidth (i.e., traffic capacity) of each edge is identical,
the maximal edge load, $L_{\texttt{max}}$, is the key factor
determining the traffic condition. Actually, the traffic congestion
will occur when $L_{\texttt{max}}$ exceeds the bandwidth. Therefore,
given a limited bandwidth, the smaller $L_{\texttt{max}}$
corresponds to higher transportation capacity. Analogously, in the
previous studies \cite{Guimera2002,Yan2006}, the maximal node load
is usually used to quantify the system's performance: the smaller
the maximal node load, the higher the transport capacity. In this
paper, we use edge load instead of node load because in the real
systems, such as the Internet and the highway, the congestion
usually happens along the edges, not at the nodes \cite{Hu2007}.

\begin{figure}
\scalebox{0.8}[0.8]{\includegraphics{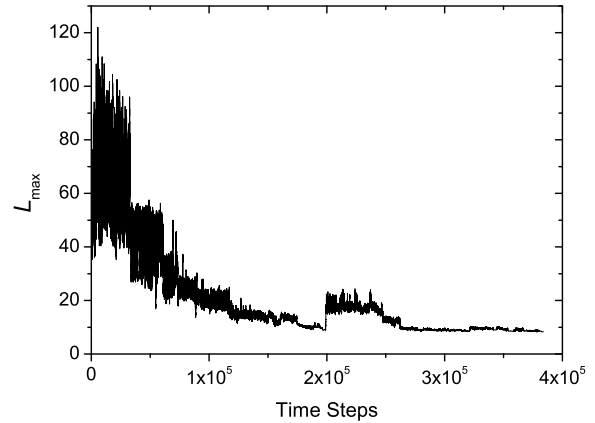}} \caption{The
objective function, $L_{\texttt{max}}$, \emph{vs}. system time, $t$,
in the optimizing process of the SA algorithm. This figure
illustrates a typical result on a BA network of size $N=1000$,
average degree $\langle k\rangle=6$. The number of suppliers is set
as $M=10$.}
\end{figure}

Given a network structure and the number of suppliers, we aim at
finding out the optimal configuration of suppliers (i.e., the
locations of suppliers) making $L_{\texttt{max}}$ as small as
possible. This is an optimization problem (indeed, an NP hard
problem) with $L_{\texttt{max}}$ being the objective function, and
the algorithm presented in this paper (see below) can be directly
extended to the case with maximal node load being the objective
function. In addition, since many real transportation networks have
heterogeneous degree distribution (see the examples shown in Refs.
\cite{Newman2006,Caldarelli2007}), we use scale-free networks to
mimic their topologies.

\section{Algorithm}
In a supply-demand network of $N$ nodes and $M$ suppliers, there are
in total $\left(\begin{array}{c} N \\ M \end{array} \right)$
different configurations for suppliers' locations. Finding the
optimal solution by evaluating all the possible configurations is
infeasible when $N\gg M \gg 1$. The optimization of a system with
many degrees of freedom with respect to a certain objective function
is a frequently encountered task in physics and beyond. One special
class of algorithms used for finding the high-quality solutions to
those NP-hard optimization problems is the so-called nature inspired
algorithms, including \emph{simulated annealing} (SA)
\cite{Kirkpatrick1983,Aarts1989}, \emph{genetic algorithms} (GA)
\cite{Holland1975,Goldberg1989}, \emph{genetic programming} (GP)
\cite{Banzhaf1998}, \emph{extremal optimization} (EO)
\cite{Boettcher2001,Zhou2005}, and so on. Here we adopt the SA
algorithm, whose procedure is as follows.

(i) Randomly choose an initial configuration, denoted by $S^0$.
Calculate its maximal edge load, $L_{\texttt{max}}^0$, and set the
best solution as: $S^{\texttt{best}}=S^0$ and
$L_{\texttt{max}}^{\texttt{best}}=L_{\texttt{max}}^0$. Set the
system time as $t=1$.

(ii) Randomly pick one supplier from the configuration $S^{t-1}$,
and change its location randomly, denote this new configuration as
$S^t$. Calculate its maximal edge load, $L_{\texttt{max}}^t$.

(iii) If $L_{\texttt{max}}^t< L_{\texttt{max}}^{\texttt{best}}$,
then set $S^{\texttt{best}}=S^t$ and
$L_{\texttt{max}}^{\texttt{best}}=L_{\texttt{max}}^t$. If
$L_{\texttt{max}}^t\leq L_{\texttt{max}}^{t-1}$, we accept the
current configuration, that is, set $t\leftarrow t+1$ and repeat
(ii). Otherwise, if $L_{\texttt{max}}^t> L_{\texttt{max}}^{t-1}$,
the current configuration is accepted with probability
$\texttt{e}^{-\Delta/T}$, where $T$ is a temperature-like parameter
and $\Delta=L_{\texttt{max}}^t- L_{\texttt{max}}^{t-1}$. When a
configuration is rejected, the algorithm directly goes back to (ii)
and keeps the system time $t$ unchanged.

To obtain the high-quality solution, one shall repeat the step (ii)
as long as desired. In this paper, we terminate the algorithm if the
variance of $L_{\texttt{max}}^t$ in the latest $10^4$ time steps is
smaller than a threshold $10^{-6}$. Note that, one time step
corresponds to one implementation of step (ii), which is different
from the system time $t$. The parameter $T$ is crucial for the
algorithmic efficiency. According to the Matropolis's Guidance
\cite{Kirkpatrick1984}, in the initial stage, the accepting
probability of a new configuration should be close to 1. Therefore,
we first choose a relatively low temperature $T_0$, and numerically
calculate the corresponding accepting probability, resulted from a
random change of one supplier's location in a completely random
configuration. The temperature is doubled until the accepting
probability reaches a threshold quantile 0.50. During the searching
process, the temperature should slowly decrease \cite{Aarts1989},
here we adopt the simplest method, that is, we set $T\leftarrow
\alpha T$ after every $Q$ time steps, where the parameter $\alpha$
is 0.90 and the period is set as $Q=0.1NM$.

\begin{table}
\caption{Comparison of the maximal edge load obtained by DTA, BTA,
GM and SA. The underlying networks are BA networks with $N=1000$ and
$\langle k\rangle=6$, and all the data are obtained by averaging
over 100 network configurations. }
\begin{center}
\begin{tabular} {ccccc}
  \hline \hline
   M / Algorithm     & DTA  &  BTA  &  GM & SA \\
   \hline
   5    & 14.73  &  14.98  &  13.32 & 12.37 \\
   10   & 8.25  &  8.92  &  7.17 & 6.31 \\
   \hline \hline
    \end{tabular}
\end{center}
\end{table}

\begin{figure}
\scalebox{0.8}[0.8]{\includegraphics{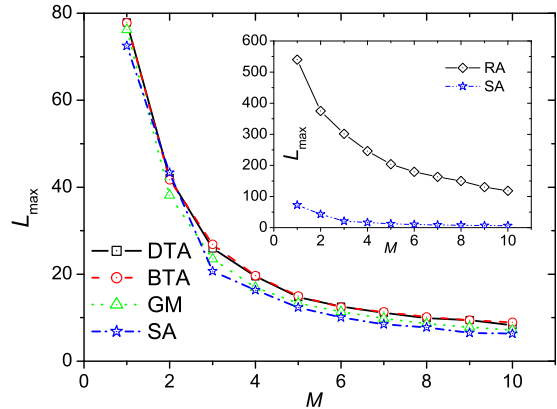}} \caption{(Color
online) Algorithmic performance for BA networks. The main plot shows
a comparison among DTA, BTA, GM and SA, while the inset reports a
comparison between RA and SA. The number of suppliers, $M$, varies
from 1 to 10, while the network size $N=1000$ and the average degree
$\langle k\rangle=6$ are fixed. All the data points are obtained by
averaging over 100 network configurations.}
\end{figure}

For comparison, we also implement some other algorithms. A brief
introduction is as follows. \emph{Random Allocation} (RA) -- The
locations of suppliers are selected completely randomly.
\emph{Degree Target Algorithm} (DTA) -- The suppliers are set as the
$M$ nodes with highest degrees. \emph{Betweenness Target Algorithm}
(BTA) -- The suppliers are set as the $M$ nodes with highest
betweennesses (see Refs. \cite{Newman2001,Zhou2006} for the
definition and calculation of node betweenness). \emph{Greedy
Method} (GM) -- First, we consider the case with only one supplier,
and find out the optimal location of this supplier that minimizes
the corresponding $L_{\texttt{max}}$. Then, we add one supplier and
find out its optimal location under the condition that the location
of the firstly added supplier is fixed. Repeating this operation,
that is, at the $k$th step, we add the $k$th supplier and find out
its optimal location subject to minimal $L_{\texttt{max}}$ under the
condition that the locations of former $k-1$ suppliers are fixed.
This algorithm is terminated when $M$ suppliers are added already.

\begin{figure}
\scalebox{0.3}[0.3]{\includegraphics{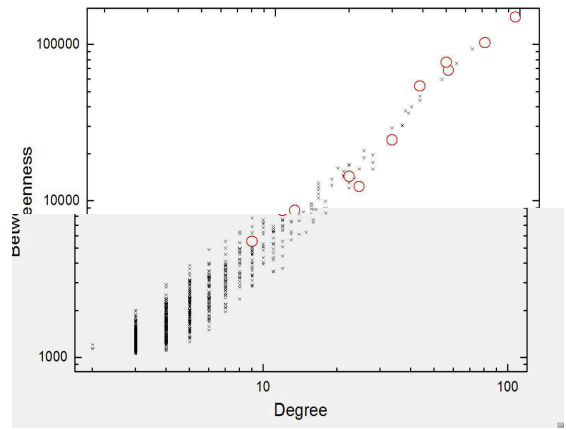}} \caption{(Color
online) Scatter plot of betweenness \emph{vs}. degree in a BA
network with $N=1000$ and $\langle k\rangle=6$. Each small black
fork represents a node. These 10 red circles denote the selected
suppliers by SA. The smallest degree of suppliers is 9, and the
second smallest one is 12.}
\end{figure}

\section{Results}
In this paper, all the numerical simulations are implemented based
on the Barab\'asi-Albert (BA) model \cite{Barabasi1999}, which is
one of the minimal models reproducing the heterogenous structure of
real-world networks. Figure 2 reports a typical optimizing process,
during which the objective function, $L_{\texttt{max}}$, fluctuates
strongly in the early stage and approaches to a relatively stable
value lately. The proposed SA can reduce the objective function,
$L_{\texttt{max}}$, by more than 10 times compared with its initial
value corresponding to a random selection of suppliers. We implement
SA in larger BA networks ($N=1000$) for different $M$ from 1 to 10,
and take the average over 100 independent network configurations. As
shown in the inset of Fig. 3, SA performs much better than RA. We
also compare SA with some mentioned algorithms, DTA, BTA and GM, and
the results have demonstrated that SA performs best. We report two
examples, $M=5$ and $M=10$, in Table 1. The improvement is in
general about 10\%. Note that, although SA performs the best, it
spends the longest running time. Actually, the time complexity obeys
the inequality $O(SA)>O(GM)>O(BTA)>O(DTA)$. Since GM performs not so
bad, it is a strong candidate especially for huge-size networks, and
GM might be a considerable tradeoff of time complexity and accuracy
of solution.

Note that, although BA model has successfully captured the degree
heterogeneity of real networks, it lacks some other important
structural properties, such as the community structure
\cite{Girvan2002} and rich-club phenomenon \cite{Zhou2004}. DTA
might perform worse if the network has strongly community structure
or presents the rich-club phenomenon. The reason is a good algorithm
should prefer to allocate suppliers to different communities rather
than putting them together in a community containing many
very-large-degree nodes, and if the very-large-degree nodes are
closely connected to form a rich club, selecting them as a whole is
of low efficiency since the increasing suppliers cannot
substantially reduce the average distance from customers to
suppliers. As a start point, we here only discuss simulation results
on BA networks, and leave the investigations of algorithmic
performance on more complicated topologies as an open issue.

The DTA and BTA have almost the same performance and give out very
similar selections of suppliers, for in BA networks betweenness and
degree are very strongly correlated \cite{Goh2001,Barthelemy2003}.
To provide insights of the solution by SA, in Fig. 4, we give a
scatter plot of betweenness versus degree, and mark by red those
selected suppliers. Though SA also prefers large-degree
(large-betweenness) nodes, the selected suppliers are remarkably
different from those by DTA or BTA, actually, moderate-degree
(moderate-betweenness) nodes also have chance to be selected by SA.
In most cases, only the top-40\% large-degree nodes have the chance
to be selected, therefore we can restrict the candidates of
suppliers in those 40\% nodes. We have check this restriction in BA
networks with $N=1000$ and $\langle k\rangle=6$, which gives out
equivalently good solution while requires about 10 times shorter CPU
time.

\section{Conclusion and discussion}
In this paper, we proposed a generic model of transport in
supply-demand network, which is consisted of suppliers (supply
nodes) and customers (demand nodes). Accordingly, a measure of edge
load is given, under the assumption that every customer only
requires service from the nearest supplier. In such a network with
heterogenous degree distribution, its transport capacity is very
sensitive to the locations of supply nodes. We therefore design a
simulated annealing algorithm to find out the near optimal
configuration of supply nodes, which remarkably enhances the
transport capacity, and outperforms the degree target algorithm, the
betweenness target algorithm, and the greedy method. This work
provides a start point for systematically analyzing and optimizing
transport dynamics on supply-demand networks. Even though the model
and algorithm are simple, we get some non-trivial result, that is,
simply picking up those nodes of highest degrees is not the optimal
method, actually, some moderate-degree nodes also have chance to be
selected as suppliers.

In our model, every customer requires the same amount of resource,
which is not in accordance with \emph{the elephants and mice
phenomenon} \cite{Papagiannaki2002} found in the real Internet,
where a small fraction of flows contribute to most of the traffic.
Corresponding to the current model, a flow stands for the resource
transported from a supplier to a customer, and thus each flow has
the same size although the one passing longer paths contributes more
to the total load. In addition, the proposed algorithm does not
fully take into account and make use of the topological features. We
have already mentioned in the last section that the mesoscopic
structure, such as communities and the rich club, may highly
influence the solutions. Those structural information should be
extracted prior to the optimizing algorithm, and be embedded in the
algorithmic procedure in some way to improve the efficiency and/or
the resulting network capacity. All those blemishes listed above can
be treated as some open problems worth of a future exploration.

To the end, we emphasize that many real systems can be better
described by the current supple-demand network model, instead of the
oversimple assumption \cite{Yan2006} that every node simultaneously
plays the roles of supplier and customer. We have already mentioned
some examples, such as power grid \cite{Liu1984,Bai2005} and supply
chain networks \cite{Cheng2006,Goh2007}, another typical example is
the software supporting systems in the Internet, where a system
usually has set up several servers in different locations, and users
from everywhere can ask for downloading of some softwares. The
locations of those servers play the crucial role in determining the
efficiency and capacity the software supporting system.

This study also provides some complementary information for relevant
phenomena in disparate systems. For example, social scientists have
studies how to design who should be integrators in a given social
communication networks to better solve problems, and they have found
that people having extensive relations (i.e., of very large degrees)
may not be the suitable information integrators, instead, the
highest efficient structure makes the distance of all nodes from the
obvious integrator the shortest \cite{Borgatti2009}, which is, to
some extent, in accordance with what we observed in this work. In
addition, empirical studies show that the public service facilities
are not just located in the place of the most dense population, but
somehow more uniformly distributed to make the total travel distance
between people and facilities shorter . As a final remark, we noted
that a very recent work has considered of the network-based
transport with multiple sources and sinks \cite{Carmi2008}, which
shows different yet relevant motivation to the current work.

\begin{acknowledgments}
This work is supported by Hong Kong Baptist University and the Hong
Kong Research Grants Council, the National Natural Science
Foundation of China under Grant No. 10635040, and the National Basic
Research Program of China (973 Program No.2006CB705500).
\end{acknowledgments}

\end{document}